\documentclass[fleqn,usenatbib]{mnras}

\usepackage{newtxtext,newtxmath}
\usepackage[T1]{fontenc}

\DeclareRobustCommand{\VAN}[3]{#2}
\let\VANthebibliography\thebibliography
\def\thebibliography{\DeclareRobustCommand{\VAN}[3]{##3}\VANthebibliography}

\usepackage{graphicx}	% Including figure files
\usepackage{amsmath}	% Advanced maths commands
\title[Magnetic flux decay in solar active regions]{Statistical analysis of the total magnetic flux decay rate in solar active regions}

\author[A. A. Plotnikov et al.]{
Andrei A. Plotnikov\thanks{E-mail: plotnikov.andrey.alex@yandex.ru (AAP)},
Valentina I. Abramenko,
Alexander S. Kutsenko
\\
Crimean Astrophysical Observatory, Russian Academy of Sciences, Nauchny, 298409, Russia\\
}

\date{Accepted XXX. Received YYY; in original form ZZZ}

\pubyear{2023}

\begin{document}
\label{firstpage}
\pagerange{\pageref{firstpage}--\pageref{lastpage}}
\maketitle

% Abstract of the paper
\begin{abstract}
We used line-of-sight magnetograms acquired by the \textit{Helioseismic and Magnetic Imager} on board the \textit{Solar Dynamics Observatory} to derive the decay rate of total unsigned magnetic flux for 910 ephemeral and active regions (ARs) observed between 2010 and 2017. We found that: i) most of the ARs obey the power law dependence between the peak magnetic flux and the magnetic flux decay rate, $DR$, so that $DR\sim \Phi^{0.70}$; ii) larger ARs lose smaller fraction of their magnetic flux per unit of time than the smaller ARs; iii) there exists a cluster of ARs exhibiting significantly lower decay rate than it would follow from the power law and all of them are unipolar sunspots with total fluxes in the narrow range of ($2 - 8) \times 10^{21}$~Mx; iv) a comparison with our previous results shows that 
the emergence rate is always higher than the decay rate. The emergence rate follows a power law with a shallower slope than the slope of the decay-rate power law. The results allowed us to suggest that not only the maximum total magnetic flux determines the character of the decaying regime of the AR, some of the ARs end up as a slowly decaying unipolar sunspot; there should be certain physical mechanisms to stabilize such a sunspot.

\end{abstract}

% Select between one and six entries from the list of approved keywords.
% Don't make up new ones.

%%%%%%%%%%%%%%%%%%%%%%%%%%%%%%%%%%%%%%%%%%%%%%%%%%

%%%%%%%%%%%%%%%%% BODY OF PAPER %%%%%%%%%%%%%%%%%%

\section{Introduction}

One of the most outstanding manifestations of the solar activity is the appearance of active regions (ARs) on the solar surface, places with much stronger magnetic flux than that in the surrounding areas. In white-light images ARs appear as groups of sunspots with low intensity. These features are not static: their shape varies during their lifetime. 

A comprehensive overview of the AR's evolution was given in \citet{vanDriel}. The life-cycle of an AR can be divided into consequent phases of growth (called as the emergence phase) and disappearing (or the decay phase). The emergence phase was explored in a variety of publications \citep[e.g.][to mention a few]{Ugarte-Urra2015, Norton2017, Kutsenko2019}. At the same time, the decay phase got much less attention. As argued by \citet{Norton2017} the reason for this is a long time interval of the decay lasting for weeks: in most cases one cannot observe the entire process since the sunspot group rotates off the limb. Usually individual ARs exist from several days up to several weeks. According to the Gnevyshev-Waldmeier rule \citep{Gnevyshev1938,Waldmeier1955}, the lifetime of a sunspot group is proportional to the maximal area of the group
\begin{equation} \label{Time-area}
T = b A_0,
\end{equation}
where $T$ is the time interval between the sunspot group's appearance and disappearance, $A_0$ is the maximal area reached by the sunspot group, and $b$ is a constant.

Through the decades, various models based on different ideas were suggested to explain the decay of the magnetic flux in ARs, for example, the self-similar sunspot model \citep{Gokhale1972}, the turbulent diffusion model \citep{Meyer1974}, the turbulent erosion model \citep{Petrovay}. The second and the third models are based on a hypothesis that the turbulence in the solar plasma plays a major role in the dissipation of the magnetic flux tube forming an AR. The difference between the diffusion model and the erosion model is in the treatment of the processes inside the tube. In the turbulent diffusion model, the key role is attributed to the turbulent diffusion of magnetic elements inside the tube, whereas the turbulent erosion model suggests that diffusion is mainly frozen inside the sunspot due to the strong magnetic field, and the outer turbulence gnaws a border of the sunspot \citep{Petrovay}.

The erosion model results in the parabolic area versus time dependence:
\begin{equation} \label{parabolic_law}
	A = A_0 - 2\sqrt{\pi A_0} w (t - t_0) + \pi w^2 {(t - t_0)}^2,
\end{equation}
where $w$ and $t$ stand for the spot boundary decrease rate (which is assumed to be a constant) and time, respectively. This dependence was confirmed in the statistical analysis by \citet{Petrovay1997} and \citet{Murakozy2021}. \citet{Svanda2021} suggested additional proofs for the erosion mechanism based on the morphological changes through the evolution of an AR. This makes the decay phase to be completely different from the emergence phase, which is thought to be driven by the turbulent diffusion mechanism. The turbulent erosion mechanism also implies sharp sunspot boundaries, which agrees with white-light observations of sunspots.

Sunspots are the observable manifestation of strong magnetic fields on the solar surface. Modern solar instruments allow us to use high-resolution data on the magnetic field. Therefore, the spatial distribution of the magnetic field can be used instead of the sunspot area in order to track the evolutionary changes in an AR. In this case, the total unsigned magnetic flux over the AR
\begin{equation}
\Phi = \int_{S}{ | (\vec{B} \cdot \vec{dS}) |}.
\label{flux}
\end{equation}
can be used instead of the total sunspot group area. In Eq.~\ref{flux} $\vec{B}$ is the magnetic field vector and $S$ is the area occupied by magnetic structures of the AR. 

In the framework of the turbulent erosion model \citep{Petrovay}, the current sheets formed around the sunspot can maintain the magnetic field strength inside the sunspot nearly unchanged. Adopted in this theory Gaussian-like distribution of the magnetic field inside the sunspot leads the magnetic flux to fall slower than the sunspots’ area during the decay phase. This is in accordance with the results by \citet{Li2021} who showed that the mean vertical magnetic field strength increases during the decay phase. Observations show that weak magnetic structures still exist after sunspots disappear. This means that an AR's lifetime will always be longer than that of the corresponding sunspot group.

Moving magnetic features \citep[MMFs;][]{Harvey1973} are often mentioned as a phenomenon, accompanying the decay of ARs. MMFs are described as small magnetic elements (usually less than 2 arcsec in size) detaching from a large magnetic concentration in an AR, running outside, and dissipating  during several hours. \citet{Kubo2008} showed that the magnetic flux transported by MMFs can be higher than the sunspot's losses of the magnetic flux. \citep{Imada2020} found a slight asymmetry in the magnetic flux carried by MMF from leading sunspots: approximately 5\% more magnetic flux is transported to the equator side than to the pole side, and about 3\% more magnetic flux is carried out to the East side than to the West side.

As we have already mentioned before, the analysis of the entire evolution of a large AR is obstructed by the Sun's rotation: the presence of the AR on the visible disc is shorter than its typical lifetime. \citet{Ugarte-Urra2015} overcame this obstacle by combining the UV data acquired by the \textit{Solar TErrestrial RElations Observatory} \citep[STEREO,][]{Kaiser2008} and by the \textit{Atmospheric Images Assembly} \citep[AIA;][]{Lemen2012} on board the \textit{Solar Dynamics Observatory} \citep[SDO;][]{Pesnell2012}. STEREO gives an opportunity to observe the solar surface from two different vantage points. Although the satellites have no equipment for magnetic field measurements, the UV intensity can be used as a proxy for the total unsigned magnetic flux \citep[e.g.][]{Schrijver1987}. To study the long-term AR evolution, \citet{Ugarte-Urra2015} measured the UV intensity of 9 ARs during their entire lifetime. The normalised intensity versus time profiles for all ARs exhibited similarity \citep[see fig.~1 in][]{Ugarte-Urra2015}. Consequently, it is reasonable to hypothesize that the lifetime of an AR is proportional to the peak magnetic flux of the AR, and the decay rate is constant for all ARs, regardless of their maximal flux.

Here we present a statistical analysis of the AR decay rates using a large data set of 910 ephemeral and ARs.

\section{Data and Methods}
\label{method}

SDO/HMI provides high-cadence (720 s) line-of-sight (LOS) full-disc 4096$\times$4096 pixel magnetograms with continuous coverage since 2010. The spatial resolution of the instrument is 1~arcsec with the pixel size of 0.5$\times$0.5~arcsec$^{2}$. High spatial resolution of the instrument allowed us to analyse small ephemeral regions exhibiting no signatures in white-light images.

\begin{figure*}
    \centering
    \includegraphics[width = 2\columnwidth]{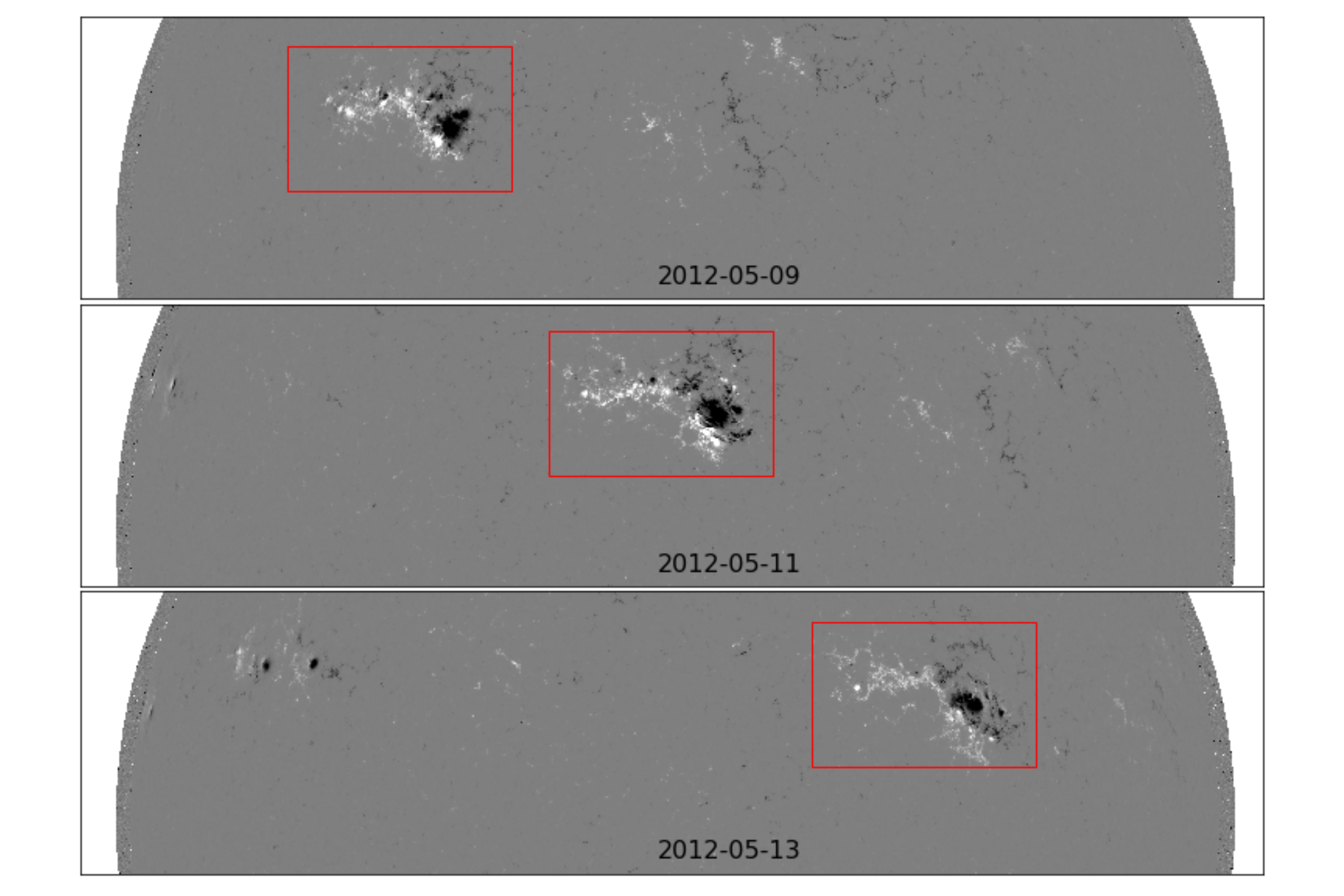}
    \caption{ Line-of-sight SDO/HMI magnetograms of the Sun, which were acquired between 2012.05.09 and 2012.05.13. Red rectangular boxes show the selected patches of NOAA AR 11476. The size of the box was set large enough to keep most of the magnetic flux of an AR inside the boundaries during the entire observations. The size of the box was kept constant in CCD coordinates.}
    \label{fig:cropping}
\end{figure*}

The magnetographic data used in this work was prepared in \citet{Kutsenko2021}. We visually analysed full-disc SDO/HMI magnetograms and manually enclosed active and ephemeral regions by a rectangular box (Fig.~\ref{fig:cropping}). The box was large enough to keep the whole AR inside the boundaries during the entire interval of observations. Thus, we visually examined the selected patches as the AR evolved. If there was a significant dispersion of the flux beyond the box boundaries, we re-selected the region and increased the box size. Consequently, the dispersed network magnetic flux that appeared during active region decay was also mostly kept within the bounding box. We selected isolated active regions in the sense that no significant portions of magnetic flux of external ARs crossed the boundaries. Each region was tracked back and forth in time in the consecutive magnetograms by a cross-correlation technique. The size of the box in CCD pixels remained unchanged. In order to minimize the uncertainties due to projection effect and noise in magnetograms, the tracking was stopped as soon as the longitude of any corner of the bounding box was equal to or exceeded 60 degrees and a thresholding was applied during the magnetic flux calculations, see below.

For unipolar active regions, the following magnetic polarity was usually dispersed over vast areas ``contaminated'' by other ARs and the magnetic connections within the region were not obvious. Hence, for these objects we selected exclusively the leading polarity of the active region.

Ephemeral regions were selected by the same manual selection. We searched for small magnetic dipoles emerging and decaying amidst quiet-Sun regions (without any pre-existing magnetic flux). We did not set requirements for ephemeral region lifetime or peak flux. In order to diminish the influence of the projection effect, we selected only ephemeral regions evolving near the disc centre. 

Thus, for each active and ephemeral region the bounded patches were cropped and stored in a data cube. In total, we prepared data cubes for 323 ephemeral and 854 active regions observed between 2010 and 2017.

Using the prepared data cubes, we calculated the total unsigned magnetic flux needed to explore the decay of ARs. Equation (\ref{flux}) can be approximated as a sum over the magnetogram:
\begin{equation}
	\tilde{\Phi} = \sum{|B_{r}|}\Delta S,
 \label{eq_flux_calc}
\end{equation}
where $B_{r}$ and $\Delta S$ stands for the radial component of the magnetic field and pixel area on the solar surface, respectively. The radial component of the magnetic field was evaluated from the observed LOS-component via the $\mu$-correction. Namely, for each pixel of the patch, we calculated the angle $\mu$ between the line-of-sight and the vector pointing from the centre of the Sun to the pixel. Both magnetic flux density and the area of the pixel were divided by the cosine of this angle. \citet{Leka2017} argued that exactly this procedure provides the best estimation of the radial magnetic field. The summation in equation~\ref{eq_flux_calc} was performed only over pixels with absolute magnetic flux density exceeding 30 Mx~cm$^{-2}$. This threshold is a fivefold noise level of SDO/HMI 720-s LOS magnetograms \citep{Liu}.

For each ephemeral and AR we derived temporal profiles of the total unsigned magnetic flux. To derive the decay rate, we need to pick out a time interval of the decay. Our requirement to the decay time interval are as follows: 

\begin{enumerate}
    \item The magnetic flux must decrease during the time interval (small oscillations of the magnetic flux can be ignored).
    \item The interval must start after the AR's emergence is finished.
    \item The interval must end either by a plateau in the total flux profile, or by a significant increase of the total flux, or by the end of observations.
\end{enumerate}

To avoid the human bias in determination of the decay segment in the magnetic flux versus time profiles, we elaborated an automatic iteration routine, which is described in details in the Appendix~\ref{appendix}. Certain profiles were rejected by the algorithm. Finally, we calculated decay rates for 241 ephemeral and 669 sunspot-containing ARs.

\begin{figure*}
    \centering
    \includegraphics{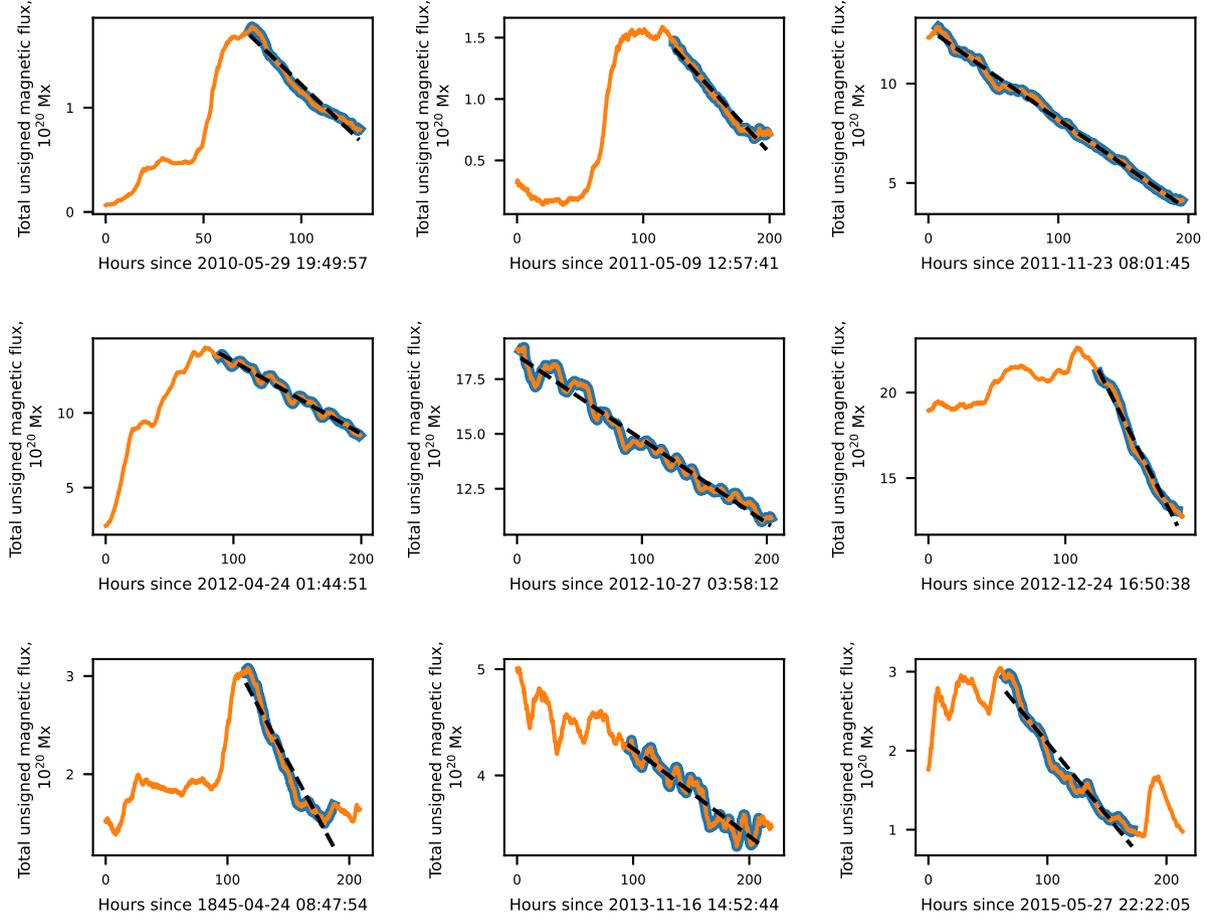}
    \caption{The total unsigned magnetic flux versus time profiles for several ARs analysed in this work (orange curves). Examples of the decay interval detection by the algorithm (see text) are shown as highlighted parts of the curves (blue).  Dashed line represents the linear fitting of the curve within the decay interval. The slope of the fitting was adopted as the decay rate, $DR$. Note the individual scales in the vertical axes.}
    \label{fig:common_graphs}
\end{figure*}

Fig.~\ref{fig:common_graphs} shows a set of examples showing the decay intervals determined using our algorithm. The decay rate, $DR$, was calculated as the slope of the linear fitting within the decay time interval. The peak magnetic flux of an AR was adopted as the maximal value of total magnetic flux along the entire temporal profile.

\begin{figure*}
    \centering
    \includegraphics{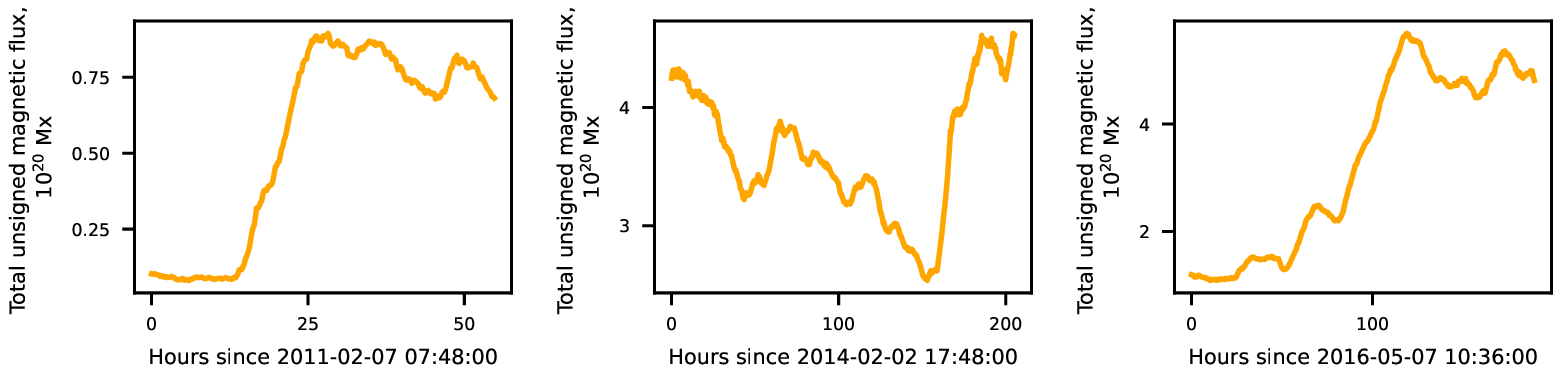}
    \caption{Examples of ARs with the total unsigned magnetic flux versus time profiles rejected by the algorithm. Note the individual scales in the vertical axes.}
    \label{fig:nodecay_graphs}
\end{figure*}

Fig.~\ref{fig:nodecay_graphs} shows three examples of magnetic flux versus time profiles rejected by the algorithm. As one can see, the algorithm failed in finding the decay interval in case of long-lasting significant emergence and in the case of jagged profiles. The total number of rejected ARs is 185 that is 22\% of all ARs in our set.

Using the  continuum intensity images acquired by SDO/HMI, all ARs in the data set were sorted out into classes of unipolar, bipolar, multipolar ARs, and ephemeral regions. 
Every selected AR was processed as the independent one. Thus, recurrent ARs were considered as independent ones at each solar rotation.

 To analyse the magnetic flux variations of opposite magnetic polarities within an AR, we calculated the total magnetic fluxes within each polarity separately:
\begin{equation}
\begin{aligned}
\Phi_{+} = \sum{B_{r}}\Delta S, \\
\Phi_{-} = \sum{-B_{r}}\Delta S.    
\end{aligned}
\label{polarity_flux}
\end{equation}

The positive and negative magnetic fluxes were calculated within the decay time interval determined by our algorithm. The decay rates for opposite polarities were also calculated by fitting the time profiles by a linear approximation within the decay interval.

In order to define the preceding polarity within an AR, we calculated the center-of-gravity for each magnetic polarity:
\begin{equation}
\begin{aligned}
CG_{x} = \sum x {B_{r}}\Delta S,   \\
CG_{y} = \sum y {B_{r}}\Delta S,
\end{aligned}
\label{cog}
\end{equation}

where $x$ and $y$ are the longitudes and latitudes of the pixels in CCD coordinates, respectively. The western polarity was assigned as the preceding one. To avoid the ambiguity, the preceding and following polarities were defined only for bipolar ARs. Similar to equation~\ref{eq_flux_calc}, all the summations in equations \ref{polarity_flux} and \ref{cog} were performed over the pixels with the magnetic flux density exceeding 30 Mx~cm$^{-2}$

In 192 ARs, we were unable to reveal the magnetic flux peak at the total magnetic flux versus time profile. The examples are shown in top-right, center-center and bottom-center panels of Fig.~\ref{fig:common_graphs}. For these ARs we adopted the maximum magnetic flux observed within the observational interval as the peak value. Hence, this set of ARs will be referred to as the ARs without the observed peak. Moreover, one should keep in mind that the observed total magnetic flux peak in the rest of ARs might be a local rather than a global maximum. In such a case, we analyse the decay rate of this newly emerged magnetic structure.

\section{Results}

\begin{figure*}
	\includegraphics[width = 2\columnwidth]{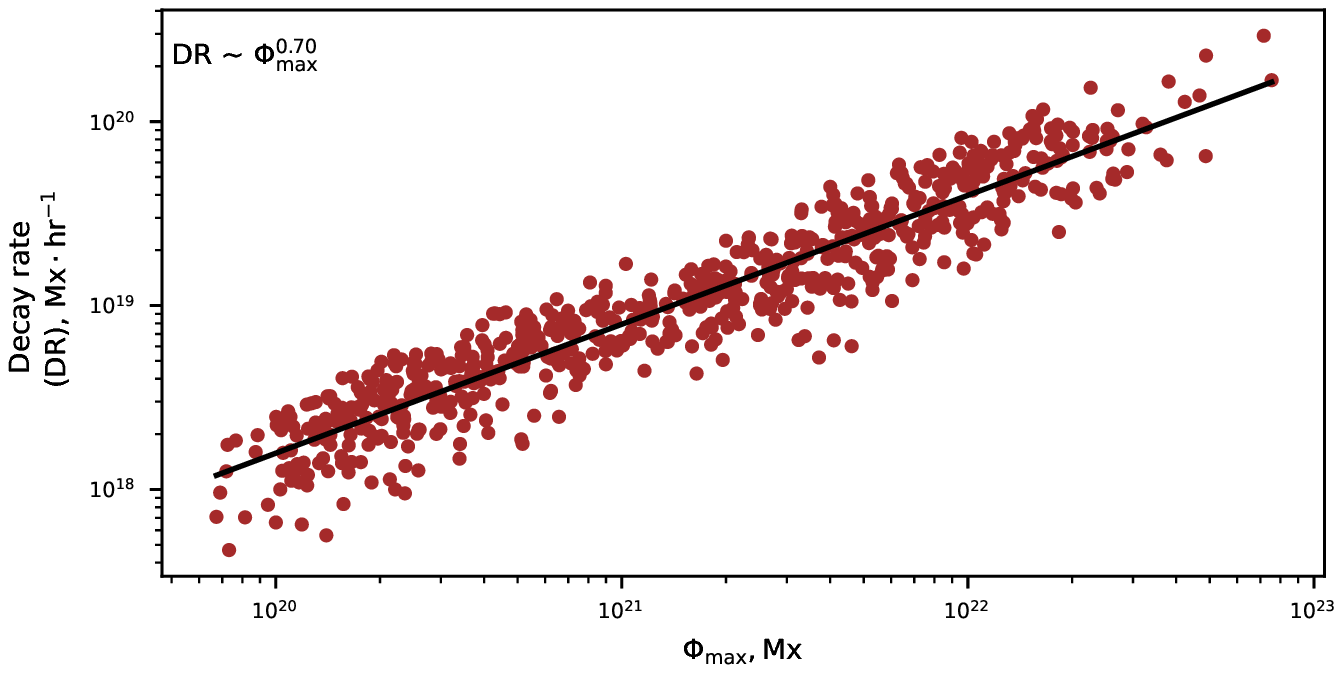}
	\caption{The magnetic flux decay rate versus the peak magnetic flux for 718 ARs with the observed total magnetic flux peak. Black line represents the linear fitting of the distribution. The power index of the fitting is $0.70 \pm 0.01$.}
	\label{fig:plot_with_peaks}
\end{figure*}

The double-logarithmic plot of the decay rate (\textit{y}-axis) versus the peak total magnetic flux (\textit{x}-axis) for 718 ARs and ephemeral regions with the observed peak magnetic flux is shown in Fig.~\ref{fig:plot_with_peaks}.

The data points are distributed along a linear fitting implying the power-law relationship between the parameters. The power index of the law is $0.70 \pm 0.01$. This means that, as a whole, the larger the AR, the higher the decay rate.

\begin{figure*}
	\includegraphics{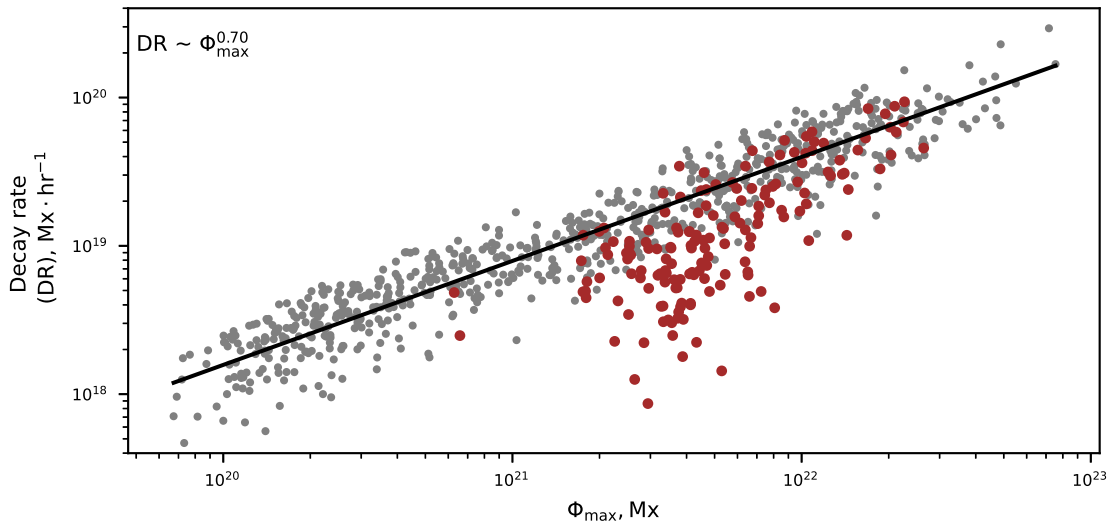}
	\caption{The magnetic flux decay rate versus the peak magnetic flux for 910 active and ephemeral regions. Both sets of ARs (with the observed peaks and without the observed peaks) are included.  Unipolar ARs are shown by red circles while all the rest of data points are shown by grey circles. Black line represents the linear fitting of the distribution shown in Fig.~\ref{fig:plot_with_peaks}.}
	\label{fig:full_ars_div}
\end{figure*}

Fig~\ref{fig:full_ars_div} shows the same plot with the addition of the set of 192 ARs without the observed peaks. The total number of ARs in this plot is 910. The black line displays the power-law relationship calculated over the data shown in Fig~\ref{fig:plot_with_peaks}. A small cluster of outstanding ARs can be revealed in the middle-bottom part of the plot. These ARs exhibit an abnormally slow decay rate (up to $\approx 10$ times slower than it could be expected from the power law). Since the total magnetic flux profile of these ARs do not exhibit a peak, the true maximum magnetic flux value is unavailable for the ARs. However, the true maximum magnetic flux value is larger (or at least not less) than the value shown in the plot. In this case, the low decay rate of these ARs is even more deviated from the values expected from the power law. Therefore, in the visual representation in Fig~\ref{fig:full_ars_div}, the red circles in the plot are expected to be shifted to the right in the $x$-direction. It makes the data points to be even farther from the power-law line.  

Our previous experience hints that there exist outstandingly long-living unipolar ARs. We explored the magnetic morphology of the ARs in the ``outstanding'' cluster in Fig~\ref{fig:full_ars_div} and revealed that these ARs were unipolar. All unipolar ARs are shown by red circles in Fig~\ref{fig:full_ars_div}. Note that not all unipolar ARs belong to the cluster: a part of unipolar ARs obey the common power-law relationship.

\begin{figure*}
	\includegraphics{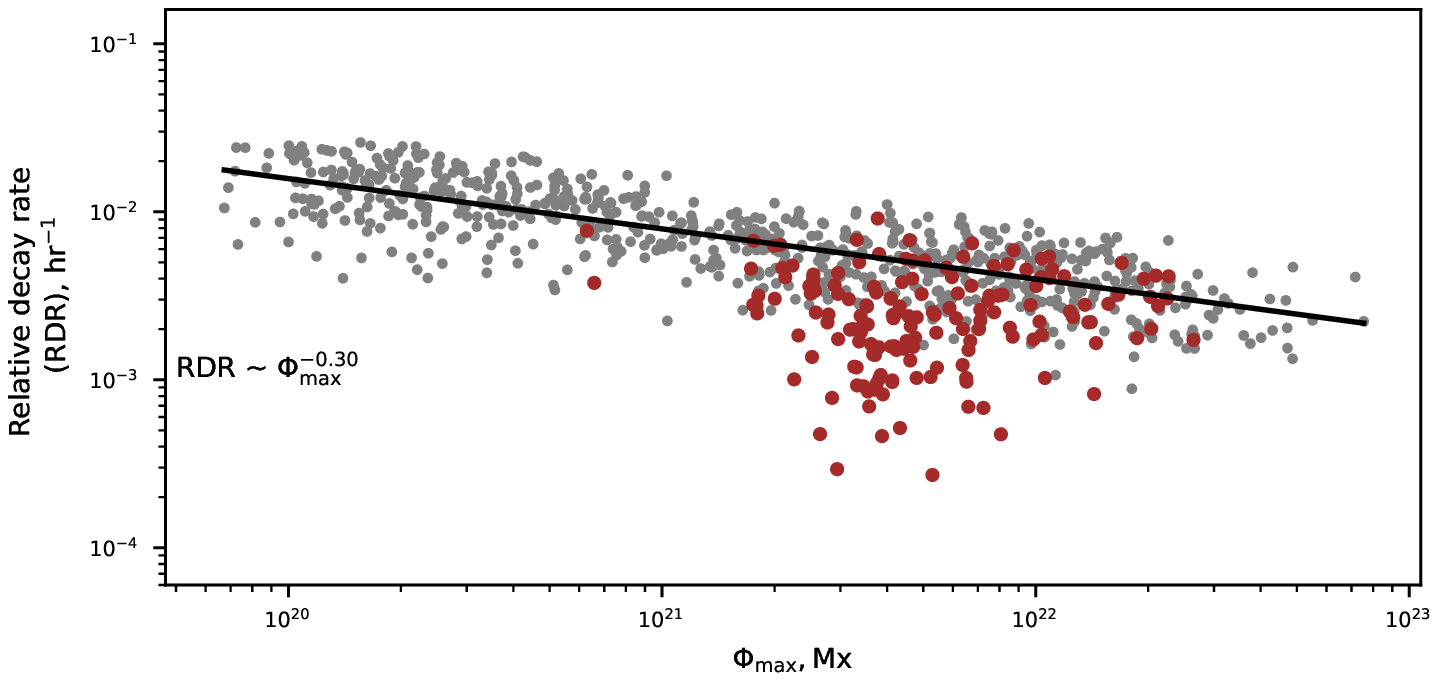}
	\caption{Same as in Fig.~\ref{fig:full_ars_div} but $y$-axis stands for the magnetic flux decay rate normalized by the peak magnetic flux. Black line represents the linear fitting of the distribution of ARs with the observed peaks only.}
	\label{fig:full_ars_div_r}
\end{figure*}

Fig.~\ref{fig:full_ars_div_r} shows the relative decay rate versus the peak magnetic flux. The relative decay rate, $RDR$, was calculated as the ratio of the decay rate to the peak magnetic flux. In other words, this value shows a fraction of the peak magnetic flux lost by an AR during a unit time (an hour). The linear fitting is derived for the set of ARs with the observed peak only. The figure shows that most of ARs satisfy the power-law with the power index of $-0.30 \pm 0.01$. The negative index implies that small ARs lose their magnetic flux faster as compared to larger ones. For example, ephemeral regions with the peak flux of $10^{20}$~Mx tend to lose more than 10\% of their magnetic flux per hour, whereas the largest ARs lose only about 1\% of their flux during the same time.
Fig.~\ref{fig:full_ars_div_r} also shows the cluster of outstanding long-living unipolar ARs. Some of them lose the magnetic flux extremely slow: the relative decay rate drops down to $10^{-3}$ that is more than order of magnitude lower than $RDR$ observed for bi/multipolar ARs. 

Another interesting feature of the ``outstanding'' cluster is the narrow range of the peak magnetic fluxes. The magnetic fluxes are located in the $(2 - 8) \times 10^{21}$~Mx range, whereas the magnetic fluxes for all unipolar ARs could differ by 50 times.

\begin{figure*}
	\includegraphics[width = 2\columnwidth]{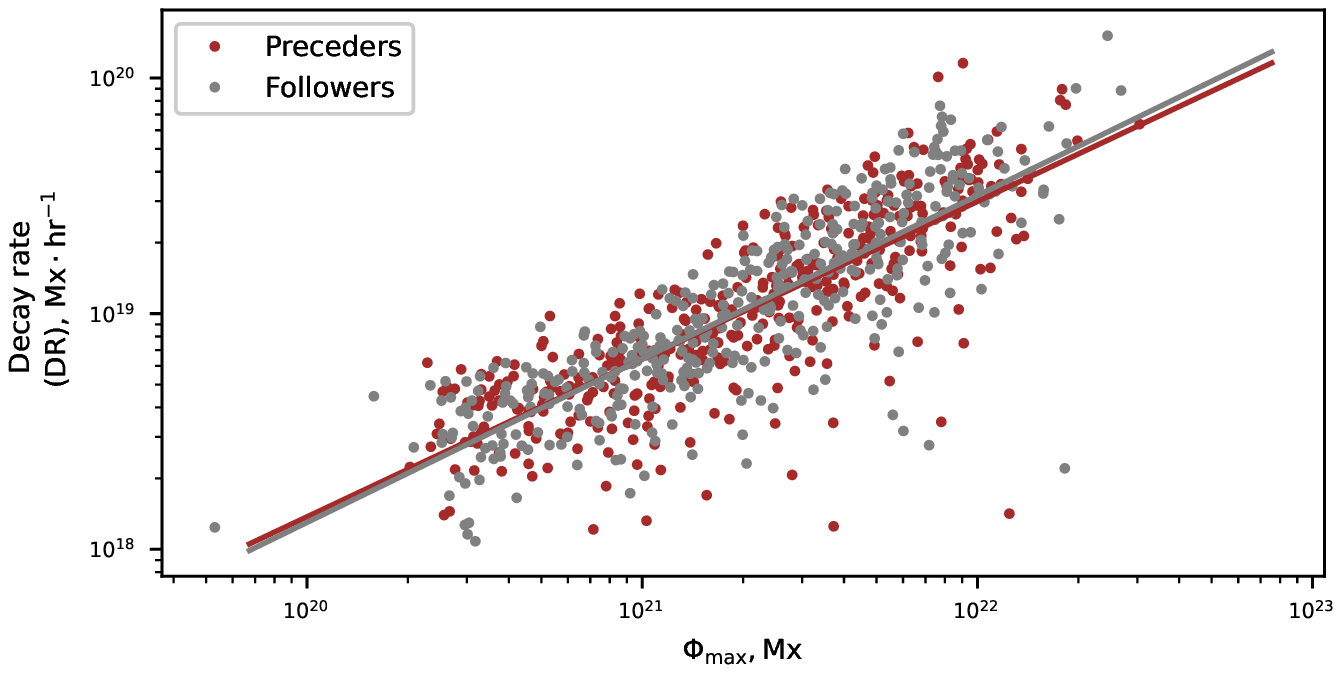}
	\caption{The magnetic flux decay rate versus the peak magnetic flux for preceding (red circles) and following (gray circles) polarities for 399 bipolar active regions. Solid lines represent linear fittings of both distributions with the same colour coding.}
	\label{fig:polar_plot}
\end{figure*}

Fig.~\ref{fig:polar_plot} shows the decay rate versus the peak magnetic flux for preceding and following polarities in 399 bipolar ARs. The fittings yield the power law indices of $0.70 \pm 0.02$ and of $0.66 \pm 0.02$ for preceding and following polarities, respectively. Seemingly, the magnetic flux losses in the preceding and following polarities obey the same power law within the uncertainties. Very close decay rates revealed for the preceding and following polarities also implies the correctness of our data reduction: the entire AR is enclosed within our bounding box and there is no significant magnetic flux loss across the boundaries.

\begin{figure*}
    \centering
    \includegraphics[width = 2\columnwidth]{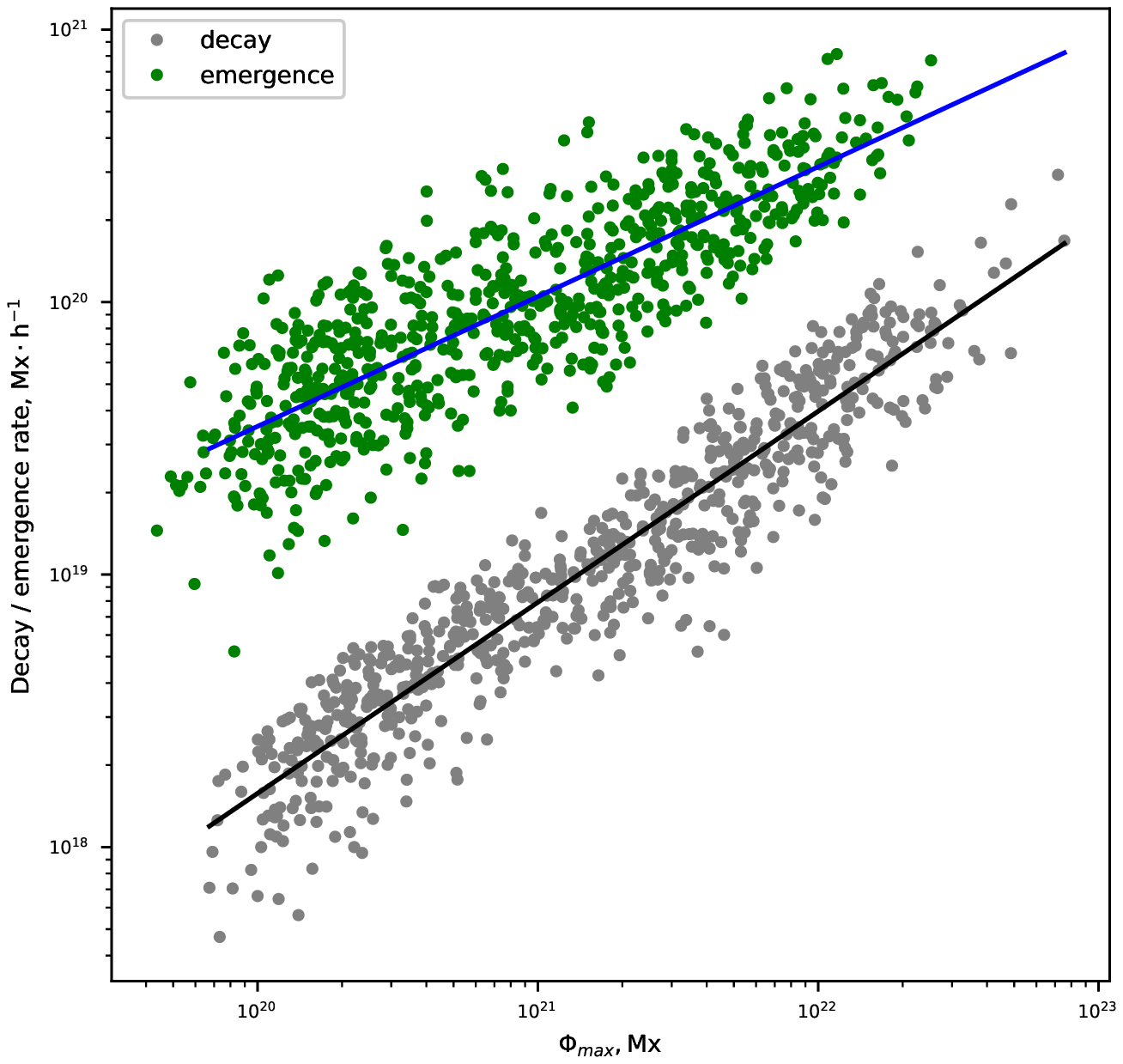}
    \caption{Magnetic flux change rate during emergence (green circles \citet[]{Kutsenko2019} plus 323 ephemeral regions added in this work) and decay (gray circles, this work) versus the peak magnetic flux. Black (blue) line shows the best linear fitting for the decaying (emerging) ephemeral and active regions.}
    \label{fig:dec_em}
\end{figure*}

We have also compared the magnetic flux change rate during emergence and decay. The flux emergence rate was measured in \citet{Kutsenko2019} for a set of 423 emerging sunspot-containing ARs by the procedures similar to those applied in this work. We supplemented the data by the flux emergence rates measured for {323} ephemeral regions from the data set compiled for this work. The results are presented in Fig.~\ref{fig:dec_em}. One can see that the emergence rate always prevails the decay rate and demonstrates the power law with more shallow slope: the power index for the emergence rate is 0.48 whereas the power index for the decay rate is 0.70. In our opinion, this difference emphasizes different physical mechanisms of magnetic flux emergence and decay.

\section{Conclusions and discussion}

In our statistical study based on SDO/HMI data acquired between 2010 and 2017, we explored the magnetic flux decay rates for 241 ephemeral and 669 active regions of different morphology. Our inferences can be summarized as follows:

\begin{enumerate}
    \item Most of ARs obey the power-law dependence between the magnetic flux decay rate and the peak total magnetic flux:
$$
    DR = 1.50 \cdot 10^{4} \Phi_{max}^{0.70}, 
$$
where the decay rate, $DR$, is in Mx~h$^{-1}$. $\Phi_{max}$ is normalized by 1.0~Mx to have unitless quantity under the exponent.
    \item Generally, larger ARs lose a smaller fraction of their magnetic flux per unit of time as compared to smaller ones. %This fact could be considered as an analogue of the Gnevyshev-Waldmeier law (eq. \ref{Time-area}) in terms of the magnetic flux.  
    
    \item Preceding and following polarities exhibit the same power-law dependence between the magnetic flux decay rate and the peak total unsigned magnetic flux.

    \item There exists a cluster of ARs exhibiting significantly lower decay rate. The cluster consists of unipolar ARs only. The peak magnetic fluxes of ARs in the cluster vary in a narrow range of ($2 - 8) \times 10^{21}$~Mx. Not all of the unipolar ARs belong to this cluster.
    
    \item  A comparison of magnetic flux emergence and decay rates confirmed that the emergence rate always prevails the decay rate and demonstrates the power law with a more shallow slope: the power index for the emergence rate is 0.48 while the power index for the decay rate is 0.70. This inference indicates that the emergence proceeds faster than the decay and they are intrinsically different processes. 
\end{enumerate}

Our results on emergence and decay rates are quite similar to that reported by \citet{Norton2017}: they found the power-law dependencies with the slopes of 0.35 and 0.57 for emergence and decay, respectively. It should be mentioned, however, that in \citet{Norton2017} the polarity-divided flux values were used. Their finding is in favour of our suggestion that emergence and decay are intrinsically different processes: emergence is mostly defined by the sub-photospheric convection whereas decay is governed by processes in the photosphere and above, where the physical conditions are different.

\begin{figure*}
    \centering
    \includegraphics[width = 2\columnwidth]{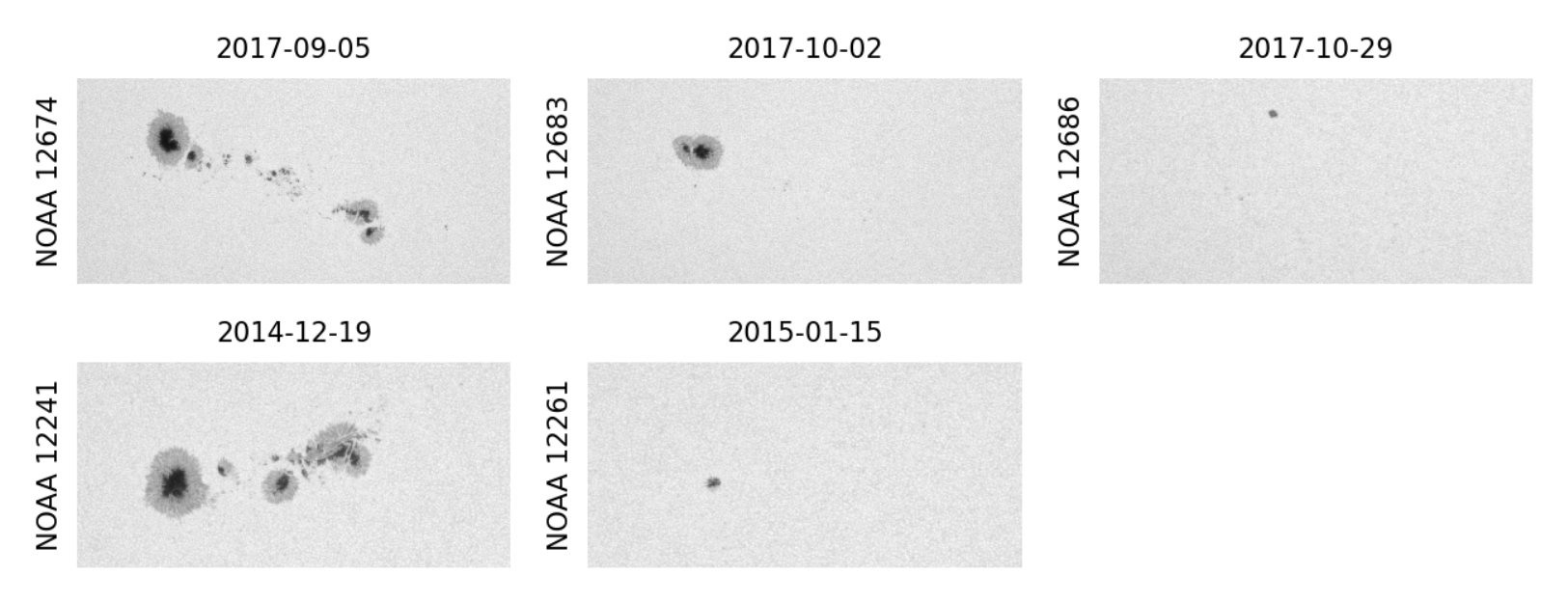}
    \caption{SDO/HMI continuum intensity images of recurrent ARs. The figure demonstrates an example of decay of two ARs with similar sunspot area. Images for the consecutive Carrington rotations are shown. The comparison shows that some of large ARs decay slowly via the stage of long-living unipolar sunspot, whereas others with the similar area decay much faster.}
    \label{fig:comparison2}
\end{figure*}

\begin{figure*}
    \centering
    \includegraphics[width = 2\columnwidth]{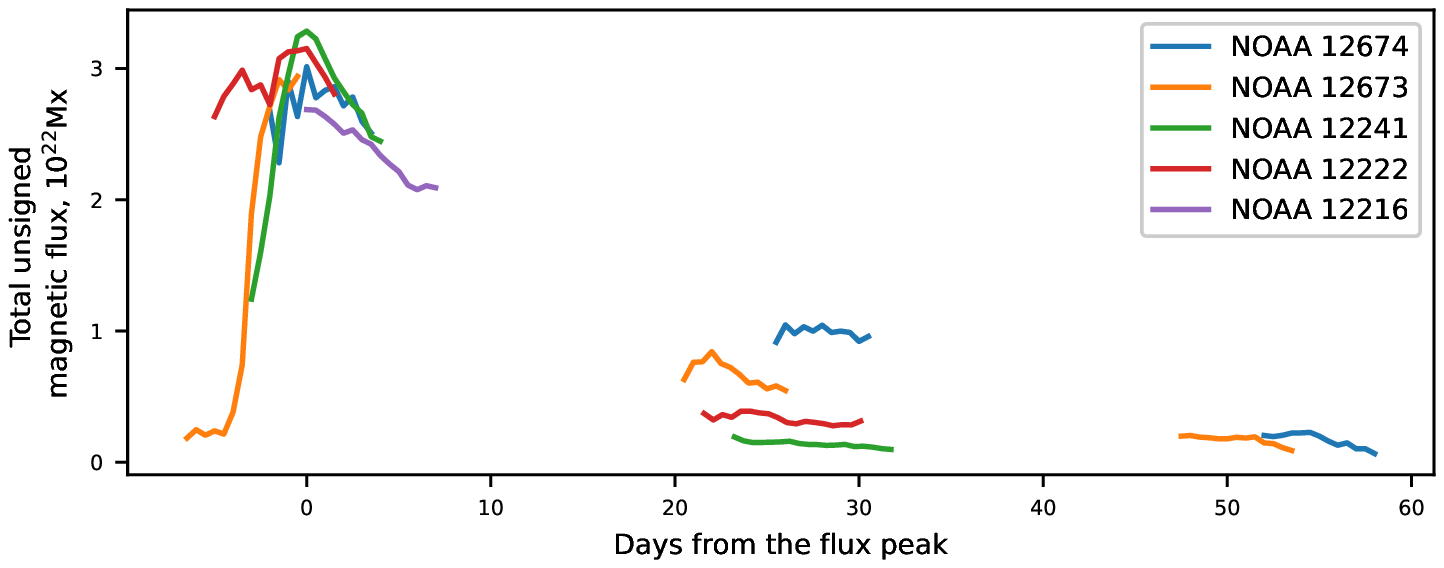}
    \caption{Time variations of the unsigned magnetic flux for five recurrent ARs. The start time $t=0$ for each AR is adopted at the time of their magnetic flux maximum. Data for three consecutive Carrington rotations are shown. NOAA AR numbers correspond to the first appearance of the AR. The colour code allows to track an AR toward the next rotation.}
    \label{fig:many_fluxes}
\end{figure*}

The revealing of a subset of extremely slow-decaying unipolar ARs implies that there exist some physical mechanism preventing regular decay in such ARs. According to \cite{Petrovay1997}, time-area relations are likely regulated by a parabolic law (Equation~ \ref{parabolic_law}), which means that area decay rate is not constant and decreases with time. Extrapolating this relations to the magnetic flux study, we could make a suggestion, that the long-life unipolar ARs could be remains of large ARs visible on the following rotation of the Sun. On the other hand, we can notice that not all of large ARs behave that way. Fig.~\ref{fig:comparison2} shows two series of recurrent ARs observed in continuum intensity by SDO/HMI. These ARs have similar peak magnetic fluxes and similar areas. At the same time, ARs' lifetimes are completely different: NOAA AR 12674 lasted at least for three rotations, while NOAA AR 12241 exhibited only a small pore on the second rotation. So the decay process must depend on more than just the peak magnetic flux. This can be further illustrated by the following experiment. 
Fig.~\ref{fig:many_fluxes} represents the unsigned magnetic flux against time profile for five recurrent ARs during three solar rotations. The ARs with close peak magnetic fluxes are selected. All of the time profiles are centered so that the peak of the magnetic flux occurs at $t=0$. Indeed, some hint of the parabolic flux decay can be tracked along the three Carrington rotations.
Nevertheless, the magnetic fluxes of the ARs widely differ during the second rotation, and only two of them (NOAA ARs 12673 and 12674) survive by the third rotation. One of them, NOAA AR 12216, has no remain that could be defined as NOAA AR even at the second rotation. Therefore, not all large ARs produce abnormally long-living sunspots, and the parabolic law of the decay could not be the only explanation for the existence of such sunspots.  

The foregoing study allows us to suggest that there should be a mechanism responsible for the appearance of the long-living ARs and their stability. Such factors as the magnetic flux imbalance, the configuration of the magnetic lines of force above the sunspots might be relevant to this phenomenon. Anyway, the results motivate further studies. 

\section*{Acknowledgements}
We are grateful to the anonymous referee whose comments helped us to improve the paper significantly. SDO is a mission for NASA’s Living With a Star (LWS) programme. The SDO/HMI data were provided by the Joint Science Operation Center (JSOC). Python programming language with NumPy \citep{numpy}, SciPy \citep{scipy} and SunPy \citep{sunpy} libraries was used for the numerical analysis. All plots were made with using of Matplotlib \citep{matplotlib} library.
%%%%%%%%%%%%%%%%%%%%%%%%%%%%%%%%%%%%%%%%%%%%%%%%%%
\section*{Data Availability}

The HMI data that support the findings of this study are available in the JSOC (http://jsoc.stanford.edu/) and can be accessed under open for all data policy. Derived data products supporting the findings of this study are available in the article and from the corresponding author (AAP) on request.

%%%%%%%%%%%%%%%%%%%% REFERENCES %%%%%%%%%%%%%%%%%%

% The best way to enter references is to use BibTeX:

\bibliographystyle{mnras}
\bibliography{decay} % if your bibtex file is called example.bib

% Alternatively you could enter them by hand, like this:
% This method is tedious and prone to error if you have lots of references
%\begin{thebibliography}{99}
%\bibitem[\protect\citeauthoryear{Author}{2012}]{Author2012}
%Author A.~N., 2013, Journal of Improbable Astronomy, 1, 1
%\bibitem[\protect\citeauthoryear{Others}{2013}]{Others2013}
%Others S., 2012, Journal of Interesting Stuff, 17, 198
%\end{thebibliography}

%%%%%%%%%%%%%%%%%%%%%%%%%%%%%%%%%%%%%%%%%%%%%%%%%%

%%%%%%%%%%%%%%%%% APPENDICES %%%%%%%%%%%%%%%%%%%%%

\appendix

\section{An algorithm for the decay interval detection }
\label{appendix}

Main features of the decay interval were described in Section~\ref{method}. Some subjectivity could be involved in the detection process. To avoid a possible ambiguity, the automatic method for the decay interval detection was elaborated.  

First, the magnetic flux versus time profile $\Phi (t)$ was smoothed with a rectangular window of 5-days width. This allows us to compensate the effects of the 24-hour HMI oscillations \citep{Liu}.

Then, we found the longest sequence $S(t)$ that satisfies the following criteria:
\begin{enumerate}
	\item	The sequence starts in $(2 m + 1$)-points local maximum of the profile (profile values are less in {\it m} previous and {\it m} following points).
    \item   The continuous fraction of the sequence over which all flux values are decreasing must lie between p and 1 times the length of the whole profile.

	\item	The maximum deviation of the data point values from the linear fitting of the sequence normalized by the maximum value along the sequence is less than $l$:
	$$
	\frac{\max_{0 < t < N} | S(t) - a t - b |}{\max_{0 < t < N} S(t)} < l, 
	$$
	where $S(t)$ is the sequence value at time $t$, $N$ is the length of the sequence, $a$ and $b$ are the linear fitting coefficients at $0 < t < N$. In other words, the sequence should not deviate considerably from the linear function.

	\item	The decrement of the linear fitting across the sequence, normalized by the entire profile $\Phi (t)$ amplitude, is larger than $r$:
	$$
	\frac{a (t_0  - t_N)}{\left( \max \Phi(t) - \min \Phi(t) \right)} > r 
	$$

The determination of the longest sequence $S(t)$ of length $N$ is performed automatically and individually for each AR in accordance with the above conditions.
	
\end{enumerate}

Then the sequence is processed to find pieces with the emergence-to-decay transitions. The starting piece of the sequence was analysed $R$ times to find the longest subsequences satisfying the following conditions:
\begin{enumerate}
	\item	The subsequence starts at the sequence’s start point;
	\item	The ratio between length of the subsequence $N_s$ and length of the whole sequence is smaller than $i$:
	$$
	\frac{N_s}{N} < i
	$$

	\item	The maximal deviation from the subsequence’s linear fitting normalized to the sequence's maximal value is less than $l$:
	$$
	\frac{\max_{0 < t < N_s} | S(t) - \hat{a} t - \hat{b} |}{\max_{0 < t < N_s} S(t)} < l, 	
	$$
	where $\hat{a}$ and $\hat{b}$ are the linear fitting coefficients for the $S(t)$ at $0 < t < N_s$;
s
	\item The ratio between the slope of the linear fitting along the subsequence, $\hat{a}$, and that of the entire sequence, $a$, is less than $k$:
	$$
	\frac{\hat{a}}{a} < k 
	$$	

	\item The sequence's ``tail'' does not decay slower than the entire sequence:
	$$
	\tilde{a} < a,
	$$
	where $\tilde{a}$ is the linear fitting coefficient for $S (t)$ at $N_s < t < N$.

\end{enumerate}

After each iteration, the found subsequence was removed from the sequence. The next iteration takes the rest of the data points as an input. 
This part of the algorithm is also performed automatically and individually for each AR.

The algorithm parameters $m, p, l, r, i, R, k$  common for the entire dataset of 910 active and ephemeral regions are listed in Table~\ref{table}. They are tweaked after several runs and the results were approved visually by each of the co-authors independently.

\begin{table}
\begin{tabular}[t]{|p{6em}|p{6em}|}
	\hline
	m & 2 \\
	\hline
	p & 0.6 \\
	\hline
	l & 0.1 \\
	\hline
	r & 0.3 \\
	\hline
	i & 0.5 \\
	\hline
	R & 2 \\
	\hline
	k & 0.5 \\
	\hline
\end{tabular}
\caption{The parameters for the automatic detection of the decay interval} \label{table}
\end{table}

%%%%%%%%%%%%%%%%%%%%%%%%%%%%%%%%%%%%%%%%%%%%%%%%%%

% Don't change these lines
\bsp	% typesetting comment
\label{lastpage}
\end{document}